\documentclass[%
 reprint,
 notitlepage,
superscriptaddress,
tightenlines,
amsmath,amssymb,aps,
]{revtex4-1}

\usepackage[T1]{fontenc} %fixes some weird stuff in the Bib

\usepackage{esint}

\usepackage{amsfonts}

\usepackage{mathrsfs}
\usepackage{empheq}
\usepackage{tcolorbox}
\usepackage{fancyhdr}
\usepackage{datetime}
\usepackage{lipsum}

\usepackage{dcolumn}% Align table columns on decimal point
\usepackage{bm}
\usepackage{color}
\usepackage{xspace}
\usepackage{amsmath}
\usepackage{nccmath}
\usepackage{soul} %\st
\usepackage[normalem]{ulem} %\sout{}
\usepackage{hyperref}

\usepackage{graphicx}
\usepackage{textcomp}
\usepackage{amsmath}
\usepackage{mathtools}
\usepackage{color}
\usepackage{xspace}
\usepackage{amsfonts}
\usepackage{mathrsfs}
\usepackage{siunitx}
\draft % marks overfull lines with a black rule on the right

\newcommand*\bq{\mathbf{Q}}
\newcommand*\bQ{\mathbf{Q}}
\newcommand*\bu{\mathbf{u}}
\newcommand*\bn{\mathbf{n}}

\newcommand*\bv{\mathbf{v}}

\newcommand*\vset{v_{\text{set}}}
\newcommand*\vstdy{v_{\text{ss}}}
\newcommand*\Kp{K_{\text{p}}}
\newcommand*\Ki{K_{\text{i}}}

\newcommand*\bI{\mathbf{I}}

\newcommand*\bE{\mathbf{E}}

\newcommand*\bx{\mathbf{x}}

\newcommand*\bOmega{\bm{\Omega}}

\begin{document}

\newcommand{\fig}{Fig. }
\newcommand{\eqn}{Eqn. }
\newcommand{\eqns}{Eqns. }
\newcommand{\jpb}[1]{{\textbf{\color{magenta} #1} }}
\newcommand{\mmn}[1]{{\textbf{\color{blue} #1} }}
\newcommand{\SF}[1]{{\textbf{\color{red} #1} }}
\newcommand{\ZD}[1]{{\textbf{\color{red} #1} }}
\newcommand{\MFH}[1]{{\textbf{\color{red} #1} }}

% Use the \preprint command to place your local institutional report number 
% on the title page in preprint mode.
% Multiple \preprint commands are allowed.
%\preprint{}

\title{Closed-loop control of active nematic flows}

% repeat the \author .. \affiliation  etc. as needed
% \email, \thanks, \homepage, \altaffiliation all apply to the current author.
% Explanatory text should go in the []'s, 
% actual e-mail address or url should go in the {}'s for \email and \homepage.
% Please use the appropriate macro for the type of information

% \affiliation command applies to all authors since the last \affiliation command. 
% The \affiliation command should follow the other information.
\author{Katsu Nishiyama}
\affiliation{The Martin Fisher School of Physics, Brandeis University, Waltham, Massachusetts 02454, USA.}
\author{John Berezney}
\affiliation{The Martin Fisher School of Physics, Brandeis University, Waltham, Massachusetts 02454, USA.}
\author{Michael M. Norton}
\affiliation{The Martin Fisher School of Physics, Brandeis University, Waltham, Massachusetts 02454, USA.}
\author{Akshit Aggarwal}
\affiliation{The Martin Fisher School of Physics, Brandeis University, Waltham, Massachusetts 02454, USA.}
\author{Saptorshi Ghosh}
\affiliation{The Martin Fisher School of Physics, Brandeis University, Waltham, Massachusetts 02454, USA.}
\author{Michael F. Hagan}
\email{hagan@brandeis.edu}
\affiliation{The Martin Fisher School of Physics, Brandeis University, Waltham, Massachusetts 02454, USA.}
\author{Zvonimir Dogic}
\email{zdogic@physics.ucsb.edu}
\affiliation{Department of Physics, University of California, Santa Barbara, California 93106, USA.}
\affiliation{Biomolecular Science and Engineering, University of California, Santa Barbara, California 93106, USA.}
\author{Seth Fraden}
\email{fraden@brandeis.edu}
\affiliation{The Martin Fisher School of Physics, Brandeis University, Waltham, Massachusetts 02454, USA.}

%Lines break automatically or can be forced with \\

\begin{abstract}

Living things enact control of non-equilibrium, dynamical structures through complex biochemical networks, accomplishing spatiotemporally-orchestrated physiological tasks such as cell division, motility, and embryogenesis. While the exact minimal mechanisms needed to replicate these behaviors using synthetic active materials are unknown, controlling the complex, often chaotic, dynamics of active materials is essential to their implementation as engineered life-like materials. Here, we demonstrate the use of external feedback control to regulate and control the spatially-averaged speed of a model active material with time-varying actuation through applied light. We systematically vary the controller parameters to analyze the steady-state flow speed and temporal fluctuations, finding the experimental results in excellent agreement with predictions from both a minimal coarse-grained model and full nematohydrodynamic simulations. Our findings demonstrate that proportional-integral control can effectively regulate the dynamics of active nematics in light of challenges posed by the constituents, such as sample aging, protein aggregation, and sample-to-sample variability. As in living things, deviations of active materials from their steady-state behavior can arise from internal processes and we quantify the important consequences of this coupling on the controlled behavior of the active nematic. Finally, the interaction between the controller and the intrinsic timescales of the active material can induce oscillatory behaviors in a regime of parameter space that qualitatively matches predictions from our model. This work underscores the potential of feedback control in manipulating the complex dynamics of active matter, paving the way for more sophisticated control strategies in the design of responsive, life-like materials.
\end{abstract}
\maketitle

\section{Introduction}

\begin{figure}    \includegraphics[width=\columnwidth]{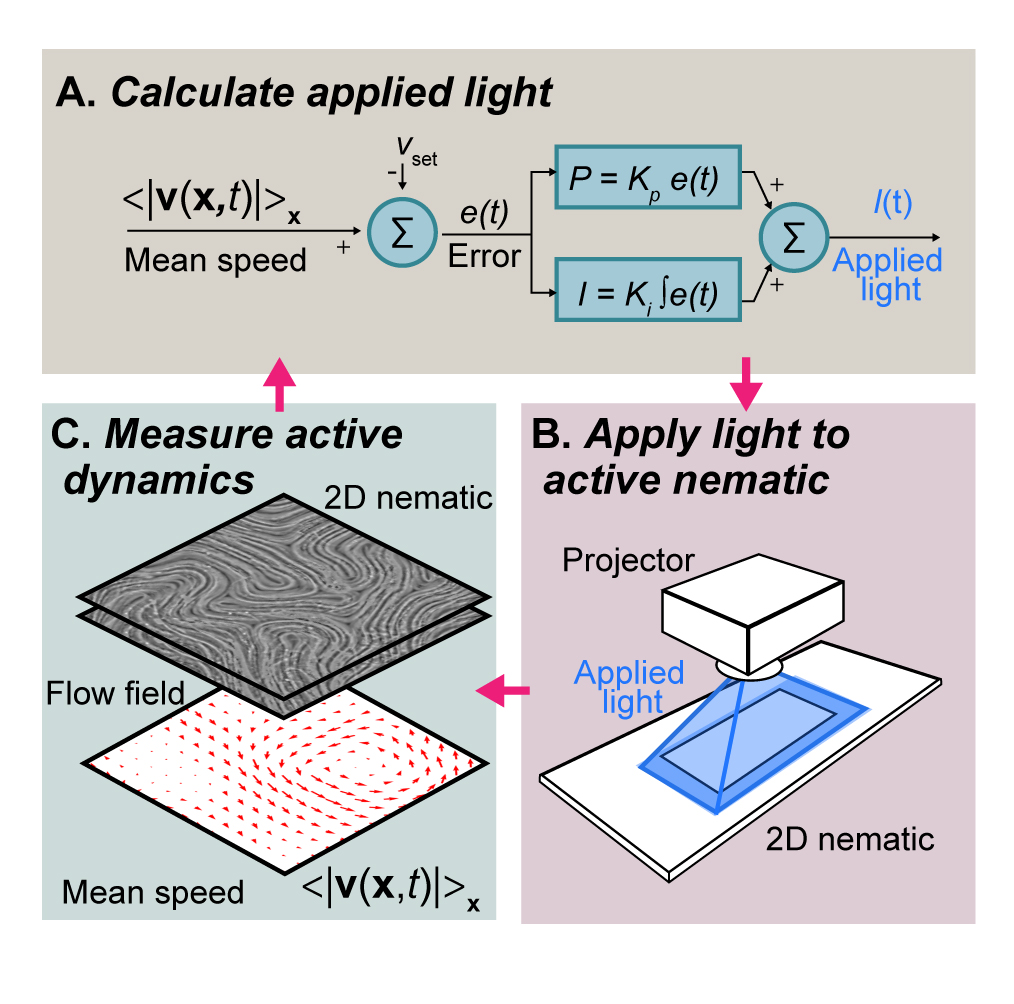}
    \caption{\textbf{Integration of hardware, wetware and software into control system} 
        (A.) We introduce control through the time-varying application of uniform light to a microtubule-based active nematic which is driven by light-sensitive kinesin motors. 
        (B.) The system state is measured through video microscopy. Optical flow analysis generates a vector field of the flow, which is reduced to a scalar magnitude. 
        (C.) A standard proportional-integral control algorithm is applied to adjust the applied light.
        }
    \label{fig:overview}
\end{figure}

Force-generating molecular motors collectively generate active stresses that drive cells away from equilibrium, thus enabling diverse life-sustaining processes such as cell division, motility, and development \cite{sedzinski2011polar, naganathan2014active,streichan2018global}. Inspired by the alluring properties found in living matter, efforts have focused on recreating life-like synthetic materials with the development of applications as a long term goal. Reconstituting materials from the bottom-up using purified cytoskeletal components is a promising route toward this goal and has revealed diverse self-organized dynamics ranging from aster formation, bulk contractions, flocking transitions, active nematic liquid crystals, spontaneously flowing active fluids, etc \cite{nedelec1997self,tabony2002biological, schaller2010polar,sanchez2012spontaneous, murrell2012f,alvarado2013molecular,foster2015active, berezney2022extensile,livne2024self}. However, despite the spectacular nature of these results, achieving a desired behavior or useful function with these systems is challenging because such self-organized far-from-equilibrium materials tend to be unstable, exhibiting chaotic dynamics and/or spontaneous switching among multiple coherent states \cite{tan2019topological, Wagner2021, wagner2023exploring}. Moreover, the protein constituents of cytoskeletal active matter are exceedingly fragile, and tend to aggregate, fragment, denature, and stick to surfaces. The dynamics of active fluids assembled from such delicate building blocks frequently reflect the nature of their constituents, exhibiting drifts that prevent obtaining long-term steady-state active dynamics. In contrast, biological processes are remarkably robust against external and internal perturbations, despite being driven by the same cast of molecular actors that are susceptible to the same forces of degradation and damage. The crucial difference enabling this robustness is that biological cells have developed intricate feedback controls that allow them to sense their current state and then adjust the force production patterns to drive toward and stabilize a particular functional dynamic \cite{steffen2017signalling}.

Given the lack of experimental active matter systems with programmable internal feedback control, we developed a system with external control, with the long term objective of replacing external control with a set of chemicals that perform a similar function. Here, we deploy a minimally complex proportional-integral (PI) controller that manipulates the spatially-averaged speed, $v=\langle|\bv\left(\bx,t\right)|\rangle_\bx$, of a light-activated microtubule nematic \cite{ross2019, zarei2023,lemma2023}. Our framework combines microscopy with machine vision \cite{tran2024} to recursively and autonomously measure system dynamics and update the applied control according to the model-free control law,
\begin{equation}
    I = \Kp \left(\vset-v\right) + \Ki\int_0^t{ \left(\vset-v\right) dt'},
    \label{eq:PIcontrol}
\end{equation} 
where $I$ is the applied intensity, and $K$ are free controller tuning parameters. While PI controllers are ubiquitous because of their simplicity to implement, they are successful only when correctly tuned. The dynamics that emerge when controller behavior interacts with the system's intrinsic behaviors can be nontrivial. For example, PI controllers can induce oscillatory behaviors even when the systems they regulate are otherwise over-damped. Controlling active systems introduces further nuances because fluctuations, which one usually attempts to minimize in passive systems, are an intrinsic feature of active fluids.

Experimentally, a variety of strategies have emerged for shaping the liquid crystalline order of active nematics and, more generally, the flows of active suspensions. These range from shaping the domain geometry
\cite{Wu2017, Wioland2013, Lushi2014, Hardouin2019, hardouin2022,Opathalage2019, Uplap2023, Fily2014, Fily2015, Fazli2021, Iyer2023, memarian2024}, viscosity or anisotropy of the underlying layer \cite{guillamat2016, guillamat2017, thijssen2021}, and spatially-varying ATP concentration \cite{bate2022}. In the case of active nematics, specifically, ad hoc dynamic interventions such as the application of externally applied flows \cite{Rivas2020} and  spatiotemporally programmed activity through applied light \cite{Zhang2019, lemma2023, zarei2023, repula2024} have demonstrated the immense potential of steering active fluids through applied fields. Spatiotemporal light control is a powerful tool for shaping the patterns of active colloids \cite{Aubret2018, Palacci2013} and microswimmers as well\cite{Volpe2011, Liebchen2018, Huang2020, Winkler2020, Villa2019, Qi2022, Katuri2017}.

A variety of theoretical strategies that determine how to actuate these various control fields in principled ways have been recently explored. At the level of hydrodynamics, model-based methods such as optimal control \cite{Norton2020, Shankar2022, sinigaglia2024, Ghosh2024} and physics-informed methods that reduce the dimensionality of the control space \cite{Shankar2024} have demonstrated immense potential. At the discrete level, optimal control can also yield inputs that guide individual particle trajectories \cite{Schneider2019, Liebchen2019}. Machine learning frameworks like reinforcement learning have been deployed \emph{in silco} \cite{Suri2024, Rotskoff2024, Falk2021} and offer an alternative method for generating control policies that don't rely on preexisting models.

However, these strategies, no matter how well-conceived, will suffer from model mismatch, unknown parameters, and the degradation of the biomolecular constituents over time, impeding their implementation. A time-tested approach that addresses such limitations is to continuously remeasure and recalculate control inputs \cite{Bechhoefer2021}. Thus, in order to join theory and experiment, feedback control is an indispensable feature that has heretofore been unaddressed in the context of active continua. 

To that end, we study here the simplest implementation of control feedback by embedding an active nematic within a Proportional-Integral control loop that drives the system toward a prescribed, spatially averaged speed. We understand the impact of the controller on system dynamics by systematically varying the proportional and integral gain values and quantifying the resulting mean speed and fluctuations. We compare these features of the experiment to theoretical predictions of a minimal linear model and full nematohydrodynamics simulation, which both include motor binding dynamics. Our observations are largely in agreement with these predictions. For proportional-only control, we observe systematic deviations from the setpoint, a well-known phenomenon called as "droop". Our measurements of droop are in excellent agreement across theory and simulation without fitting parameters.  By examining the fluctuations at steady state, we observe critically damped and underdamped (oscillatory) regimes, whose timescales vary with gain values.  Notably, we find that accounting for motor binding dynamics is essential for recapitulating observations of oscillations for proportional-only control in a simple model, which guides us towards a minimal coarse-grained theoretical framework that includes one  empirically characteristic timescale for active hydrodynamic fluctuations and one for motor binding.

\section{Methods}
\subsection{Experimental}
We developed a system which integrates a 2D active nematic (AN) with both hardware and software to implement feedback control. The microtubule-based active nematic is actuated by light-sensitive kinesin motor clusters. Control of the system is realized through varying the applied light emitted from a projector focused directly onto the sample stage (Fig. \ref{fig:overview}A). The control loop's software uses the Pycromanager library to coordinate the various hardware components of the control system. Our routine collects fluorescence images of the AN, calculates the mean speed using a machine learning-based optical flow algorithm (Fig. \ref{fig:overview}B), and updates the uniformly applied light intensity, $I$, of the projector according to equation \eqn \ref{eq:PIcontrol} (Fig. \ref{fig:overview}C).

Samples are enclosed in rectangular chambers composed of glass whose bottom wall is coated with Rain-X to create a hydrophobic surface and whose top surface is grafted with polyacrylamide to create a hydrophilic surface. These chambers are about \SI{1}{\milli\meter} tall and about \SI{18}{\milli\meter} by \SI{5}{\milli\meter} in the sample plane. Fluorinated oil is flowed into the chamber. Following this, aqueous solutions containing MTs, kinesin, polyethylene glycol depletant, ATP, oxygen scavengers and ATP-regeneration components are flushed through, creating an oil layer beneath the aqueous sample. The details of the active sample are provided in the SI\cite{ESI} and are similar to previous preparations. Our implementation differs from previous realizations because the volume of the chamber is increased  $\sim$10-fold to increase the capacity of phosphoenolpyruvate (PEP), which is the ATP-precursor in our ATP-regenerative enzymatic system~\cite{zarei2023}. The sample is then centrifuged for 3 minutes to quickly concentrate the MT bundles at the oil-water interface and placed on the fluorescence microscope within the control system. High intensity light is applied to activate the bundles and pack them into a two-dimensional film. The active nematic is imaged with fluorescence microscopy and the control system is then activated. Before feedback control is recorded, experimental calibration is performed to allow for interpretation of data and comparison to our theoretical predictions. We measure the minimum flow speed of the active nematic in the absence of applied light, $v_0$, as well as the linear response as light is increased, $\beta$.

\subsection{Theory}
To frame our observations of both the steady state mean behavior and its fluctuations, we developed a minimal linear ordinary differential equation model that describes the coarse-grained dynamics of the spatially-averaged speed $v$ and concentration of dimerized motors $m$ such that:
\begin{align}
    \dot{v} &= -v/\tau_v + \gamma_v m +\eta\left(t\right), \label{eqn:secondtorder_velocity}\\
   \dot{m} &= \left(m_0-m\right)/\tau_m + \gamma_m I, \label{eqn:secondtorder_motor}
\end{align}
A key feature of this model is that the control input $I\left(t\right)$ does not act directly on the velocity, but through the intermediate dynamics of motor activation.  This model assumes an infinite pool of motors available for dimerization. While more complex models of light-activated materials built from cytoskeletal components have been recently explored \cite{yang2024}, here we focus on the simplest case so as to minimize the number of parameters in the model and facilitate analytical calculations. In brief, \eqn\ref{eqn:secondtorder_velocity} considers that the average speed $v$ can change in 3 ways. The term $-v/\tau_v$ represents the intrinsic relaxation dynamics of the active nematic, $\gamma_v m$ represents activity driven acceleration and $\eta\left(t\right)$ represents activity driven noise.  \eqn\ref{eqn:secondtorder_motor} considers that the motor concentration that drives the active nematic can vary for two reasons. The first, represented by $\left(m_0-m\right)/\tau_m$  describes the dynamics of bound motors in the absence of light, with $m_0$ the amount of permanently bound motors, while $\gamma_m I$ describes how the motor concentration varies in the presence of light.

The response of the system is governed by timescales for hydrodynamic and motor dynamics, $\tau_{v,m}$. The latter was measured previously $\tau_m\sim\mathcal{O}\left(10\right)\text{[s]}$ \cite{zarei2023}, and the former we gather from our experiments here, $\tau_v\sim\mathcal{O}\left(100-1000\right)\text{[s]}$. The model is driven by white noise $\eta\left(t\right)$. We note that when modeling the control of a system, noise often represents exogenous forcing. In our case, the dominant contributor of noise is intrinsic to the active system itself, and, as such, we will subsequently let the amplitude of the noise scale with activity. 

Finally, we introduce the proportionality coefficients $\gamma_{m,v}$. Their exact values are unimportant for the steady state behavior of the system (average behavior and fluctuations) because, as we will see shortly, they only depend on a group of parameters coinciding with the experimentally measurable constant $\beta$, introduced below.

We first consider features predicted by this model of the limiting case of proportional-only control, e.g. $\Ki=0$. For proportional-only control, we solve \eqns\ref{eqn:secondtorder_velocity},\ref{eqn:secondtorder_motor} for the steady-state velocity with  $I=\Kp \left(\vset-v\right)$ and find
\begin{equation}
  \frac{\vstdy}{\vset}= \frac{\left(\beta \Kp+v_0/\vset\right)}{\left(1+\beta \Kp \right)}, 
  \label{eq:droop}
\end{equation}
 where $\beta$ is the measured proportionality between applied light and velocity at steady state $\vstdy\sim \beta I_{\text{ss}}$ and is defined in terms of model coefficients $\beta\equiv \tau_v \tau_m \gamma_v \gamma_m$ (See SI for a typical calibration curve used to estimate $\beta$ from experiment \cite{ESI}). $v_0\equiv m_0 \tau_m \gamma_v$ is the measured dark speed. To facilitate comparison between experiment for which $v_0\neq0$ and simulation, $v_0=0$, we rearrange \eqn\ref{eq:droop} and define a dimensionless steady state speed accounting for the offset
\begin{equation}
    \vstdy^*\equiv \frac{\vstdy}{\vset} - \frac{v_0}{\vset}\frac{1}{1+\Kp^*}=\frac{\Kp^*}{1+\Kp^*},
    \label{eq:droop_offset}
\end{equation}
where $\Kp^*=\beta\Kp$ is the dimensionless gain.

In order to compare our measurements of the fluctuations of the average speed, we examine the frequency response of the model \eqns\ref{eq:PIcontrol},\ref{eqn:secondtorder_velocity},\ref{eqn:secondtorder_motor} subject to white noise $\eta$. As mentioned above, we posit that the dominant source of fluctuations arises from the active hydrodynamic flows, so we let the noise drive the system at the level of the velocity dynamics, \eqn\ref{eqn:secondtorder_velocity}. To account for the impact of droop on these intrinsic fluctuations, we let the amplitude scale with the square root of the steady state mean speed (\eqn\ref{eq:droop_offset}), such that $\eta\sim \sqrt{\vstdy^*}\sim \sqrt{\Kp^*/\left(1+\Kp^*\right)}$. Taking the Fourier transform of all functions $F\left(\omega\right)=\int_{-\infty}^{\infty} \text{d}t\ e^{-  i  \omega t} f\left(t\right)$ and solving the resulting linear system for the transformed velocity $V\left(\omega\right)$ gives

\begin{equation}
    V\left(\omega\right) = H\frac{i  \omega \tau_v\left(1+i\omega\tau_m\right)}    {\Ki^*+i\omega\Kp^*+i\omega\left(1+i\omega\tau_m\right)\left(1+i\omega\tau_v\right)},
    \label{eq:PSD}
\end{equation}
where, $H$ is the amplitude of the white noise, and, as in previous calculations, the unknown factors $\gamma_{v,m}$ are absorbed into the experimentally measurable factor $\beta$, and $\Kp^*,\Ki^*=\beta\Kp,\beta\Ki$ are the scaled gains. Plotting $|V|^2$ yields predictions as a function of parameters $\{\Kp^*,\Ki^*,\tau_{v},\tau_m\}$. Finally, we identify the criterion for oscillations by solving the characteristic equation for the eigenvalues of \eqns\ref{eq:PIcontrol},\ref{eqn:secondtorder_velocity},\ref{eqn:secondtorder_motor}
\begin{equation}
\lambda^3+
\lambda^2\left(\frac{1}{\tau_m}+\frac{1}{\tau_v}\right)+
\lambda\frac{1+\Kp^*}{\tau_m\tau_v}+
\Ki^*\frac{1}{\tau_m\tau_v}
=0,
    \label{eq:eigen}
\end{equation}
and finding conditions where $\text{Im}\left(\lambda\right) > 0$.

\subsection{Numerical}
We compare experimental observations to a standard model of nematohydrodynamics subject to PI control and aforementioned motor binding dynamics. Nematic order is represented by the traceless and symmetric second order tensor $\bQ=s\left[\bn\otimes\bn-\bI/2\right]$ where $\bn$ is the nematic orientation and $s=\sqrt{2\bQ:\bQ}$ is the degree of order. Momentum conservation,  in the limit of low inertia and vanishing passive liquid crystalline stresses, and incompressibility govern the fluid flow $\bu$ and hydrostatic pressure $p$ such that
\begin{align}
\partial_t\bQ+\bv\cdot\nabla\bQ+\left[\bOmega,\bq\right]-\lambda\bE&=\left(1-s^2\right)\bq+K\nabla^2\bQ, \\
\nabla^2\bv-\nabla p - \alpha\nabla\cdot\bq -\xi \bv&=0,\\
\nabla\cdot\bv=0,
\end{align}
where $2\Omega_{ij}=\partial_{j}v_i-\partial_{i}v_j$ and $2 E_{ij}=\partial_{j}v_i+\partial_{i}v_j$ are, respectively, the anti-symmetric and symmetric parts of the flow field gradient, $[\cdots]$ is the commutator. In our study, we fix the strength of substrate friction $\xi = 0.01$, nematic elasticity $K = 4$, and flow alignment $\lambda=1$. 

The feedback control input $I\left(t\right)$ does not modulate active stress directly, instead it drives the dimerization of motors through the same dynamics applied to our coarse-grained model \eqn\ref{eqn:secondtorder_motor}. Since we apply control to the entire domain, we can ignore convection and diffusion processes in $m$. We fix $\tau_m=0.1$ in our simulations.

We solve the coupled system of PDEs, ODE, and control law using the spectral solver cuPSS \cite{caballero2024} in a domain size of $128 \times 128$ with $ 128 \times 128$ numbers of spatial modes.

\begin{figure}
    \includegraphics[width=\columnwidth]{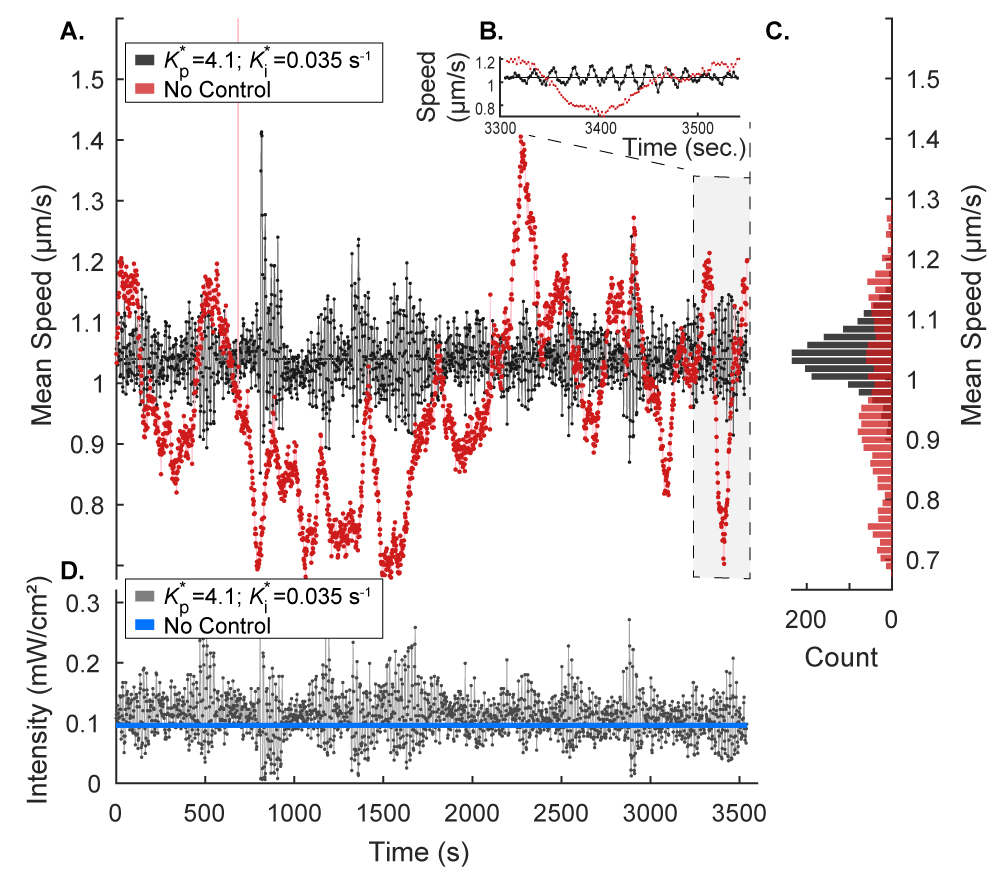}
    \caption{\textbf{Feedback control dynamics adjusts the applied light to regulate active nematic speed}
        (A.) Trajectory of PI controlled nematic (black circles) and uncontrolled nematic (red circles) with target speed of \SI{1.04}{\micro\meter\per\second}.
        (B.) Time series demonstrating speed oscillations in PI controlled nematic
        (C.) Histogram of the active nematic speed for PI controlled (black) and uncontrolled nematic (red). Mean speed $\pm$ standard distribution are \SI{1.04 \pm 0.05}{\micro\meter\per\second} for PI controlled nematic and  \SI{0.97 \pm 0.23}{\micro\meter\per\second} for uncontrolled nematic.
        (D.) Applied light over time for controlled (gray) and uncontrolled (blue) systems. The controller algorithmically determines the appropriate applied light to follow the desired speed profile (solid line, black) }
    \label{fig:control_demo}
\end{figure}

\section{Results}

We found that a PI controller regulates the steady-state behavior of the AN against both self-organized disturbances intrinsic to its active nature and those arising from aging and aggregation of its biomolecular components. We systematically varied $\Ki$ and $\Kp$ to demonstrate the role these control gain values play in achieving control. Figure \ref{fig:control_demo} illustrates regulation of an AN under a single set of PI control gains in comparison with an uncontrolled nematic of similar average speed over one hour. In comparison with the uncontrolled nematic (red), the PI controlled material (black) is centered within 0.1\% of the set speed and its standard deviation is significantly reduced (5.0\% vs. 24\% of mean speed). To accomplish this regulation, the applied light is modulated by the PI controller algorithm using input from the realtime measurement of the AN flow, \fig\ref{fig:overview}. We use an empirically derived relationship between applied light and flow speed to rationally examine the role of varied PI control gains. The measured flow speed increases nearly linearly with applied light from \SIrange{0}{0.15}{\milli\watt\per\centi\meter\squared} and saturates beyond this \cite{ESI}. In the linear regime $\vstdy\sim\beta I_{\text{ss}}$; we will use the empirically measured proportionality constant $\beta$ throughout in order to compare system responses across samples. From experiment to experiment, $\beta$ deviates by about $40\%$ which leads to significant challenges for open-loop control design \cite{ESI}. In our case, these deviations do not affect the ability to control the sample. However, we  measured $\beta$ in each experiment in order to compare data across experiments. Additionally, we found the light-responsiveness of the nematic changes slowly over time due to aging and protein aggregation, see SI for details \cite{ESI}. PI control can regulate the speed despite these changes in the control parameter and is simple to implement without any model.

\begin{figure}
    \includegraphics[width=\columnwidth]{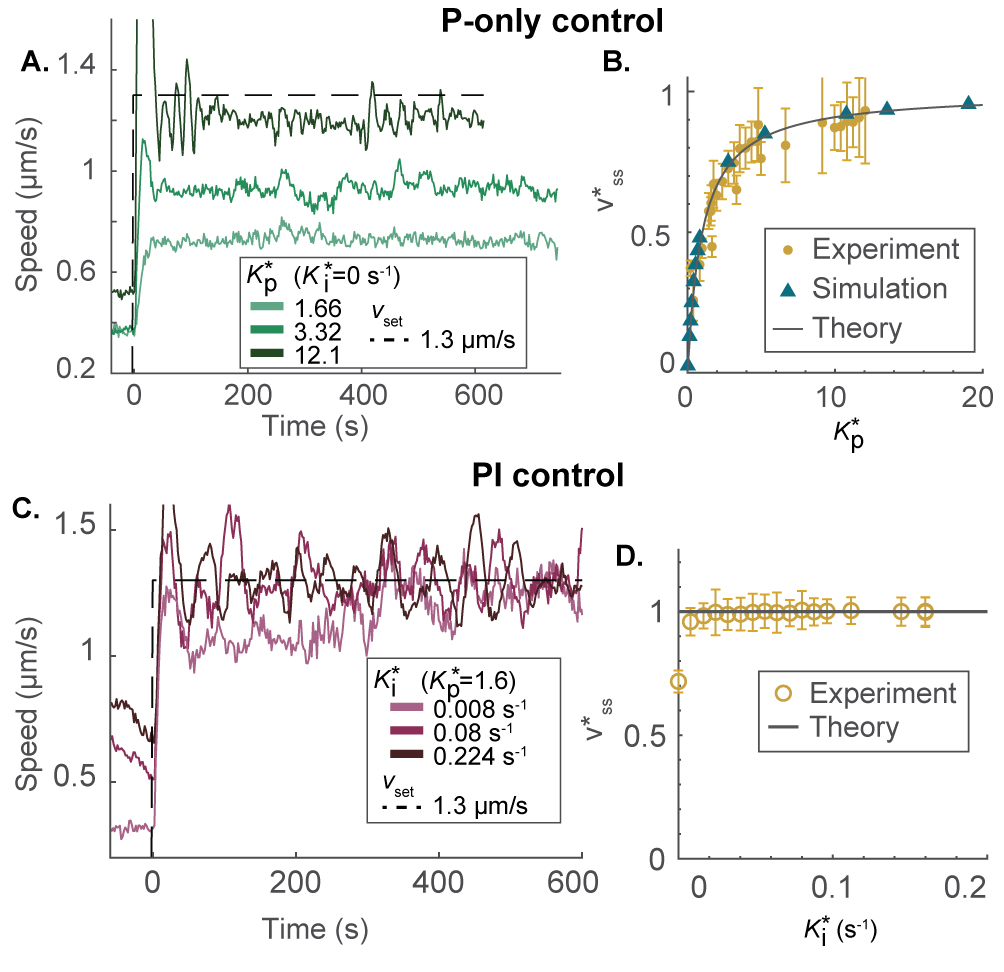}
    \caption{\textbf{ Varying PI control parameters exhibits expected behavior} 
        (A.) Response of system with varied $K_{P}$ with a set speed of \SI{1.3}{\micro\meter\per\second}. 
        (B.) Steady state droop as a function of proportional gain matches theoretical predictions.
        (C.) Response of system at set speed of \SI{1.3}{\micro\meter\per\second} with varied integral gain terms.
        (D.) Steady state droop disappears when the integral term is used.  
        }
    \label{fig:role_of_control_params}
\end{figure}

To understand the role of control parameters on dynamics, we first consider proportional-only control by systematically varying $\Kp$ with $\Ki=0$. In proportional control, the system input is proportional to the instantaneous deviation of the system from the set point. A drawback of this control law is its requirement for some steady state error, known as droop, to produce a control signal. The measured offset from the set speed is consistent with this behavior and depends on the proportional gain, \fig\ref{fig:role_of_control_params}A. Across all values of $\Kp$, we observed finite droop that decreased monotonically with increasing $\Kp$ \fig\ref{fig:role_of_control_params}B.

\begin{figure*}
    \includegraphics[width=\textwidth]{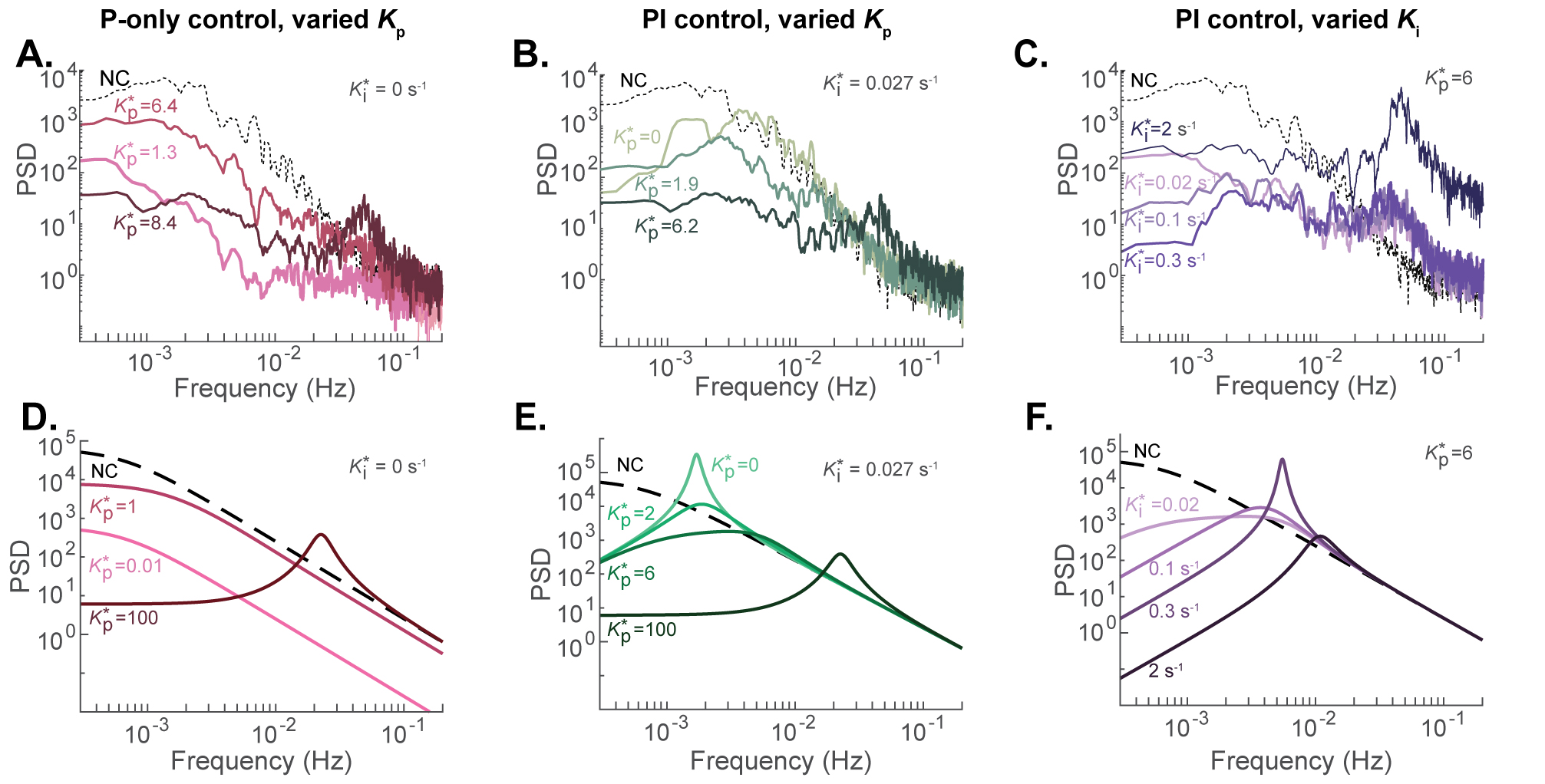}
    \caption{\textbf{Changing control gains impacts the power spectra of fluctuations in the mean speed $v$}. Top row are experimental observations and bottom row are theoretical predictions from \eqn\ref{eq:PSD}. (A,D) shows the emergence of oscillatory fluctuations for proportional-only control at large gains in both experimenta and model, (B,E) Introduces a small amount of integral gain, which results in non-monontic behavior as $\Kp^*$ is varied, and (C,F)
\label{fig:fluctuations_under_control}}
\end{figure*}

In \fig\ref{fig:role_of_control_params}, we compare the prediction for the dimensionless steady state velocity to both experiment and nematohydrodynamic simulations by plotting \eqn\ref{eq:droop_offset} against the dimensionless proportional gain $\Kp^*=\Kp\beta$ and find that the three coincide without fitting parameters. In all cases, the velocity asymptotically and monotonically approaches the set velocity with increasing gain. We emphasize bridging experiment and simulation with theory is possible because we directly measure the proportionality, $\beta$, and the speed in the absence of light, $v_0$. 

In practice, simply increasing $\Kp$ to approach the set point induces unwanted dynamics such as large overshoots, \fig\ref{fig:role_of_control_params}A.  Integral control addresses the issue of droop by accumulating the error from the setpoint over time and adjusting the control to bring the offset to zero. Any amount of $\Ki$ will drive the system to the set point but increasing the $\Ki$ increases the error accumulation rate, leading to faster controller response. When we introduce integral control to the AN system, droop is eliminated and the flow speed fluctuates around a time-average mean equal to the set speed (\fig\ref{fig:role_of_control_params}c). \fig\ref{fig:role_of_control_params}d shows this is successful for all $\Ki \neq 0$.

In the time series shown in \fig\ref{fig:role_of_control_params}a,c, we readily see fluctuations around the steady state. Error bars in \fig\ref{fig:role_of_control_params}b and d, quantify the standard deviation and indicate some systematic changes in the fluctuations as a function of control parameters. Thus, in addition to regulating the mean steady-state speed, controller parameters strongly influence  fluctuations. To further quantify them, we compute the power spectral density of the spatial average of the flow speed at steady-state. 

\fig \ref{fig:fluctuations_under_control}A-C, shows the role of proportional-only and proportional-integral control on the fluctuations of the system. Broadly, the addition of feedback control leads to a decrease in the system fluctuations compared to uncontrolled samples. However, $\Kp$ and $\Ki$ gains do not uniformly affect fluctuations across the spectra. Instead, the controller-AN system selects particular timescales and, in some combinations of controller gain, amplifies the fluctuations above the uncontrolled basis (\fig\ref{fig:fluctuations_under_control}A). 

The heterogeneous action of the controller on the timescale of fluctuations is ubiquitous and is demonstrated by the emergence of peaks in the Power Spectrum Distribution  (PSD), which represent incipient oscillatory modes. 
For example, peaks near  $f\sim\SI{5e-2}{\hertz}$ are present in each of \fig\ref{fig:fluctuations_under_control}A-C. In the supplement, we plot the corresponding autocorrelation functions $\langle v\left(t\right)v\left(t-t'\right)\rangle$ and show the switch from purely exponentially decaying correlations to oscillatory behavior. These peaks erupt in \fig\ref{fig:fluctuations_under_control}A and B when the $\Kp$ is sufficiently increased. \fig\ref{fig:fluctuations_under_control}D and E shows corresponding theoretical predictions for a selection of $\Kp^*$ values. They show that increasing $\Kp^*$ results in the emergence of a peak around $f\sim \SI{3e-2}{\hertz}$, in agreement with the experiment. The model clarifies that these coherent fluctuations are a result of the controller interacting with the motor response timescale of the active nematic. The time lag associated with motor dynamics in our model is essential for creating oscillations under proportional-only control. If, by contrast, we assumed that control acted directly on the velocity field such that $\dot{v}=-v/\tau_v+\gamma \Kp \left(\vset-v\right)$, no oscillations would be possible because the eigenvalue is necessarily real. 

The response of the system fluctuations to control is bracketed by another, longer timescale, resulting from the relaxation of the active nematohydrodynamics. This timescale is imprinted on the PSD of the uncontrolled nematic as the crossover frequency of the Lorentzian-like form (\fig\ref{fig:fluctuations_under_control}, dashed lines). Its interaction with the controller is exhibited in the experimental PSD of \fig\ref{fig:fluctuations_under_control}B. At $\Kp=0$ (given fixed $\Ki = 0.027$), a broad peak is centered at \SI{3e-3}{\hertz}. As $\Kp$ is increased, this peak erodes and, at $\Kp=6.2$, a new peak grows at \SI{5e-2}{\hertz}. This behavior, which shows a transition in the system between two distinct oscillatory states governed by the interaction of the controller with the two separate intrinsic timescales, is recapitulated in our model (\fig\ref{fig:fluctuations_under_control}E). 

In other regimes of PI control, our model predicts oscillatory behavior at timescales that are a blend between the two intrinsic timescales (\fig\ref{fig:fluctuations_under_control}F). In contrast, our experimental observations do not validate these predictions (\fig\ref{fig:fluctuations_under_control}E). Instead, we observe the gradual onset of oscillations at the motor-response timescale (\fig\ref{fig:fluctuations_under_control}C). While the behaviors in \fig\ref{fig:fluctuations_under_control}A and B are perturbations on limiting behavior, this is deep in control parameter space and may be subject to significant experimental error. In particular, our measurements of $\beta$ are subject to error, causing difficulty in exactly matching the specific regime of control behavior which governs the mixing of timescales.

Next, we focus on another key feature of the proportional-only control experiments that is captured in our model. In \fig\ref{fig:fluctuations_under_control}A, the total power varies non-monotonically with $\Kp^*$. At the lowest gain, the fluctuations are significantly reduced compared to the non-controlled case across all frequencies. Increasing the gain to $\Kp=6.4$ leads to an increase across the power spectrum. But, as the gain is increased further, fluctuations below \SI{2e-2}{\hertz} are reduced. The corresponding theoretical predictions (\fig\ref{fig:fluctuations_under_control}D) show that increasing the gain from $\Kp^*=0.01$ to $1$ indeed increases the amplitude across all frequencies as droop is reduced. Further, increasing to a larger gain $\Kp^*=100$ results in a decrease in the amplitude (and the emergence of a peak as discussed earlier).

A final feature of note in the power spectra is the non-monotonic change in the magnitude of fluctuations under $\Kp$-only control. This phenomenon arises because the controller impacts fluctuations in two ways. Firstly, even in the absence of control, fluctuations scale with the steady state applied light; as activity increases, so too do the deviations from the mean. Consequently, one expects fluctuations to be small for small $\Kp^*$ and for them to increase with gain as droop is reduced, \fig\ref{fig:role_of_control_params}B. However, since deviations from the set point are also penalized by the controller, and this penalty increases as $\Kp$ is increased, there is an eventual tradeoff between these opposing trends. In other words, there are two regimes: one at a low proportional gain, where the controlled system generates small fluctuations compared to the penalty of the $\Kp$ error term, and another, at high $\Kp^*$, where the system produces large fluctuations, which are strongly penalized and reduced. We capture these features in our predictions, \fig\ref{fig:fluctuations_under_control}D, by letting the amplitude of the systems intrinsic fluctuations scale with the $\Kp^*$-dependent steady state of the system, such that $\eta \sim \sqrt{\Kp^*/\left(1+\Kp^*\right)}$.

\begin{figure}
    \includegraphics[width=\columnwidth]{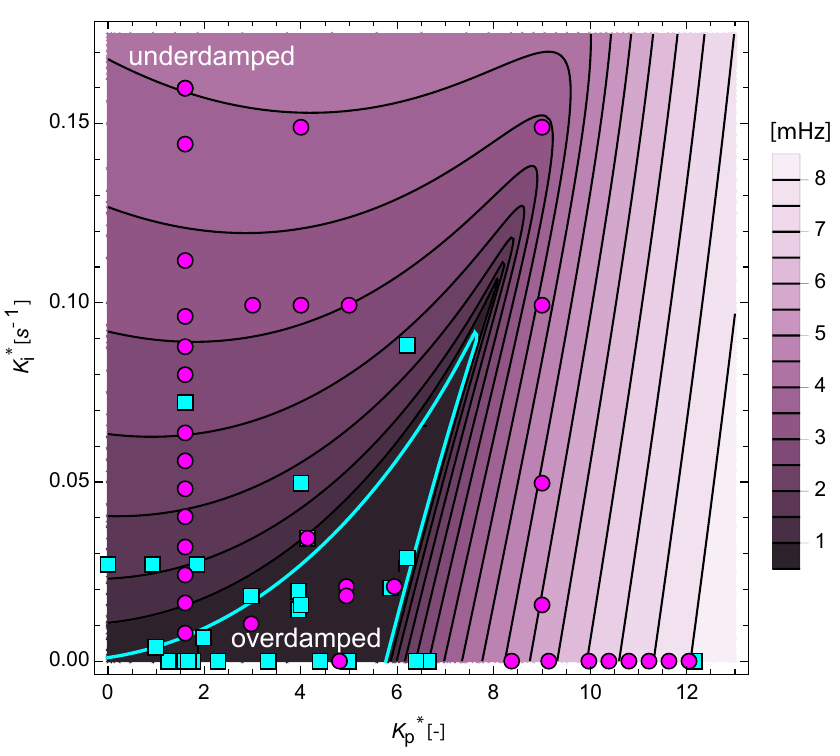}
    \caption{\textbf{Comparison between experimental and theoretically-predicted phase diagram} (using $\tau_m=10$s and $\tau_v=250$s) as a function of scaled gains $\Kp^*,\Ki^*$. Data point symbols indicate oscillatory (magenta circles) or purely exponentially decaying auto-correlation functions (cyan squares), see supplementary information for details \cite{ESI}. The cyan line demarcates the theoretical boundary between underdamped and overdamped regimes with contours indicating the frequency of oscillations.}
    \label{fig:phasediagram}
\end{figure}

We summarize our observations of the power spectra in a phase diagram \fig\ref{fig:phasediagram} in $\{\Kp^*,\Ki^*\}$, where symbols indicate the presence (circles) or lack (squares) of control-induced oscillations.  We compare to theory by examining where the eigenvalues $\lambda$ of the combined PI-nematic system, \eqn\ref{eq:eigen}, become imaginary and, in the oscillatory regime, plot contours of the frequency. The interior of the wedge region marked by the cyan contour is overdamped, $\text{Im}\left(\lambda\right)=0$, while the region beyond it supports oscillations, $\text{Im}\left(\lambda\right)>0$. Theoretical predictions corroborate the main features of our observations that the system can enter an oscillatory regime by increasing either $\Kp^*$ or $\Ki^*$ and that doing so also increases the frequency of oscillations once they are present.  We emphasize that a model accounting for the time-lag associated with motor dynamics was essential for recreating the experimental behavior in the special case $\Ki^*=0$. Theoretical predictions of frequency highlight an experimental challenge for assessing whether an experiment exhibits oscillations; near the transition (cyan), frequencies are low $\sim\SI{1e-3}{\hertz}$ and, therefore, are poorly resolved on the timescale of our experiments. We attribute further mismatch between the theory and experiment to the scaling factor $\beta$, which can drift by a factor of two over the course of experiments, as well as complexities in motor binding dynamics that our minimal model does not capture. To convey the sensitivity of the model to its parameters, we show how the predicted phase boundary changes with $\tau_{v,m}$ in the supplement \cite{ESI}.

\section{Conclusion}
Controlling the composition and physical environment of active materials and observing the consequent emergent dynamics has been a fruitful general strategy for understanding the principles organizing their behavior. The ability to apply spatiotemporal inputs enables entirely new classes of experiments that interact dynamically with active materials through responsive hardware and software.  In particular, these tools allow one to consider a range of control goals, such as stabilizing intrinsically unstable behaviors or driving the materials to altogether new regions of phase space that are otherwise inaccessible with static, spatially uniform activity. The possibilities are immense, but a key process needed to execute control tasks is feedback. In this work, by implementing a PI control algorithm, we've laid the groundwork for highly integrated experiments wherein software and hardware systems interface with active materials in real time and utilize spatiotemporally-resolved algorithms to create new behaviors. 

\begin{acknowledgments}

This work, including experiments and computation, was primarily supported by the U.S. Department of Energy, Office of Science, Office of Basic Energy Sciences under Award No. DE-SC0022291 (KN,JB,SG,AA,MFH,ZD,SF);  development of the analytical theory was supported by DOE BES No. DE-SC0022280 (MMN).

\end{acknowledgments}

\bibliographystyle{apsrev4-1}

\end{document}